\documentclass[12pt]{article}
\usepackage{cite}
\usepackage{amsmath,latexsym,epsfig}
\usepackage{amsfonts,amssymb,amscd}

\newcommand{\grgen}{{\xi}}

\usepackage{amsmath}
\usepackage{amssymb}
\usepackage{epsfig}

\begin{document}

% macros
\def\cM{{\cal{M}}}
\def\cP{{\cal{P}}}
\def\cT{{\cal{T}}}
\def\cV{{\cal{V}}}
\def\cW{{\cal{W}}}

\def\bfi{{\rm\bf i}}
\def\bfu{{\rm\bf u}}
\def\bfx{{\rm\bf x}}
\def\bfy{{\rm\bf y}}
\def\bfM{{\rm\bf M}}

\def\bbP{{\mathbb{P}}}
\def\reals{{\mathbb{R}}}
\def\comp{{\mathbb{C}}}
\def\integ{{\mathbb{Z}}}
\def\tsh{{\textstyle{\frac{1}{2}}}}
\def\tsqu{{\textstyle{\frac{1}{4}}}}
\def\tr{{\rm{trace\,}}}
\def\rintn{\int_{\reals^N}\!\!}
\def\symgp{{\mathfrak{S}}}
\def\av#1{\left\langle#1\right\rangle_{\cW_N}}
\def\efM{e^{-\tsqu\tr\bfM^2}}
\def\etM{e^{-\tsh\tr\bfM^2}}
\def\ip#1#2{\left\langle #1,#2\right\rangle}
\def\avv#1{\left\langle#1\right\rangle_{\cV_N}}
\def\Fix{{\sf{Fix}}}
\def\ipt#1#2#3{\left\langle #1,#2\right\rangle_{#3}}
\def\stat{{\sf{stat}}}
\def\ones{{\bf 1}}
\def\rel{\,{\sf rel}\,\,}
\def\mybar{\overline}

\def\proof{{\rm\bf Proof:\quad}}
%-------------

%%%%%%%%%%%%%%%%%%%%%%%%%%%%%%%%%%%%%%%%%%%%%%%%%%%%%%%%%%%%%%%%%%%%%

\begin{titlepage}

% the footnote symbols are only redefined for the title page !
\renewcommand{\thefootnote}{\alph{footnote}}
\vspace*{-3.cm}
\begin{flushright}

\end{flushright}

\vspace*{0.3in}

{\begin{center} {\Large\bf The  Goulden-Harer-Jackson matrix  model}

\end{center}}

\vspace*{.8cm} {\begin{center} {\large{\sc
               
                }}
\end{center}}
\vspace*{0cm} {\it
\begin{center}
 \vspace*{.8cm} {\begin{center} {\large{\sc
                Noureddine~Chair
                }}
\end{center}}
\vspace*{0cm} {\it
\begin{center}
 Physics Department,
 University of Jordan, Amman, Jordan 
 \begin{center} and
 
The Abdus salam International centre for Theoretical Physics, Trieste, Italy
 
     Email: n.chair@ju.edu.jo\\ \hspace{19mm}\\
\end{center}
\end{center}}
\end{center}}

\vspace*{1.5cm}

\begin{center} Abstract\end{center}  An alternative formula for the partition function of the Goulden-Harer-Jackson matrix  model is derived, in which the Penner and the orthogonal Penner partition functions are special cases of this formula. Then the free energy that computes the parametrized Euler characteristic  $\xi^s_g(\gamma)$ of the moduli spaces as yet an unidentified, for $g$ is odd, shows that the expression for  $\xi^s_g(\gamma)$ contains the orbifold Euler characteristic of the moduli space of Riemann surfaces of genus $g$, with $s$ punctures for all parameters $\gamma$. The other contributions are written as a linear combinations of Bernoulli polynomials at rational arguments . It is also shown that in the continuum limit, both the Goulden-Harer-Jackson matrix  model and  the Penner model have the same critical points.  
 \vspace*{.5cm}
\end{titlepage}

\renewcommand{\thefootnote}{\arabic{footnote}}
 Goulden, Harer and Jackson in their interesting paper\cite{Harer1} obtained an expression for the parametrized Euler characteristic  $\xi^s_g(\gamma)$, a polynomial in $\gamma^{-1}$ which gives  when specializing the parameter, $\gamma$, to  $\gamma=1$ and  $\gamma=1/2$,  the orbifold Euler characteristic of the moduli space of complex algebraic curves ( Riemann surfaces ) of genus $g$  with $s$ punctures,  real algebraic curves (non-orientable surfaces) of genus  $g$  with $s$ punctures, respectively. It was shown explicitly that  $\xi^s_g(1/2)$ for $g$ odd   coincides with  the orbifold Euler characteristic of the moduli space of complex algebraic curves \cite{Harer2},  \cite{Penner} while, if $g$ is even,  $\xi^s_g(1/2)$ corresponds  to the orbifold Euler characteristic of the moduli space of real algebraic curves, this may also be called the orthogonal Penner model  \cite{Chekhov}, \cite{Chair1}. One must say that the Penner approach  is more accessible to physicists since it uses techniques from  Feynman diagrams and random matrices \cite{Bessis}. Here, we   will give an alternative  formula for the partition function from which $\xi^s_g (\gamma)$ may be computed. The simplicity of this formula is that  the partition functions of the Penner and the orthogonal Penner models are transparent, and so  this formula may be considered as a parametrized partition function for the Goulden-Harer-Jackson model.  Having done so,  it will be shown that if $g$ is odd,  the  parametrized Euler characteristic  $\xi^s_g(\gamma)$ is a sum of two terms one of which is the orbifold Euler characteristic of the moduli space of complex algebraic curves and the other term is  written as a linear combinations of  of the Bernoulli polynomials at rational arguments. For $g$ even,  $\xi^s_g(\gamma)$ is shown to  coincide with the results obtained  previously  in \cite{Harer1}. It remains to give the geometrical interpretation of  moduli spaces whose Euler Characteristic is given by $\xi^s_g(\gamma)$, also give  the physical  meaning of the free energy that computes  $\xi^s_g(\gamma)$, up-to now, this moduli space is not yet identified. By taking the continuum limit of the present model, then, it turns out that both the Penner and the Goulden-Harer-Jackson models have the same critical points.  The partition function for the  Goulden-Harer-Jackson model may be written as
 \begin{eqnarray}
\label{p0} 
W_\gamma(N,t)=
\frac{
\displaystyle{
\int_{\reals^N} \vert \Delta(\lambda)\vert^{2\gamma}
\prod_{j=1}^N e^{-i\gamma\lambda_j/ \sqrt{t}}
e^{-\frac{\gamma}{t}\log(1-i\sqrt{t}\lambda_j)}d\lambda_{j}
}
}{
\displaystyle{
\int_{\reals^N} \vert \Delta(\lambda)\vert^{2\gamma}\prod_{j=1}^Ne^{-\gamma \sum_{i=1}^{N}\lambda_{i}^2/2}d\lambda_{j}}},
\end{eqnarray}
wherre $\Delta(\lambda)= \prod_{1\le i<j\le N}(\lambda_j-\lambda_i)$ is  the Vandermonde determinant. If we set $\gamma=1$, then, this is the Penner Model \cite{Penner}, and so this model may be considered  as a deformed Penner model, the deformed parameter being $\gamma$.
Now, the parametrized Euler characteristic was shown in\cite{Harer1} to be  connected to the partition function $W_\gamma(N,t)$, through the following expression 
\begin{eqnarray}
\label{p1}
{\grgen}^s_g (\gamma)= s!(-1)^s[x^s t^{g+s-1}]\,
\frac{1}{\gamma}\log W_\gamma(N,t),
\end{eqnarray}
where $ X[Y]$,  is a short notation for the coefficient $ X$ in the expansion of $Y $, and  $$W_\gamma(N,t) =\left(\sqrt{2\pi}\,\, e^{-\gamma/t}
\left(\frac{\gamma}{t}\right)^{\frac{\gamma}{t}-\tsh((N-1)\gamma+1)}\right)^N
\prod_{j=0}^{N-1}\frac{1}{\Gamma\left(\frac{\gamma}{t}-\gamma j\right)}.$$
is the partition function  obtained from Eq. (\ref{p0}), using the Selberg integration formula. 
 To motivate our method in obtaining  $\xi^s_g(\gamma)$,  let us first consider the case in which $\gamma=1/2$. In this case, we may use the Legendre duplication formula $ \Gamma(z)\Gamma(z+\frac{1}{2})=\frac{\sqrt{\pi}}{2^{2z-1}}\Gamma(2z)$, to show
\begin{equation}
\label{p2}
\prod_{j=0}^{N-1}\Gamma\left(\frac{1}{2t}-\frac{1}{2j}\right)=\prod_{j=0}^{N/2-1}\frac{\sqrt{\pi}}{2^{1/t-1}}\Gamma\left(\frac{1}{t}-N-(2j+1)\right),
\end{equation}
and from the identity
\begin{equation}
\label{p3}
\Gamma\left(\frac{1}{t}-N-(2j+1)\right)=\frac{t^{N-(2j+1)}\Gamma\left(\frac{1}{t}\right)}{\prod_{j=0}^{N-(2j+1)}(1-pt)}.
\end{equation}
we get
 \begin{eqnarray}
 \label{p4}
 W_{1/2}(N,t) =\left(\frac{\sqrt{2\pi t}\,\, (et)^{-1/t}}{\Gamma\left(\frac{1}{t}\right)}
\right)^{N/2}\prod_{j=0}^{N/2-1}\prod_{p=1}^{N-(2j+1)}(1-pt).
 \end{eqnarray}
By setting $\gamma=1/q$, $N=qK$, Goulden, Harer and Jackson derived the following formula
\begin{eqnarray}
\label{p5}
W_\frac{1}{q}(qK,t)=
\left(\frac{\sqrt{2\pi t}
}{ \Gamma(\frac{1}{t})\,(et)^\frac{1}{t}
}\right)^K
\frac{{\prod_{l=1}^K\prod_{j=1}^{ql}(1-jt)}
}{ {\prod_{j=1}^K(1-tqj)},
}\end{eqnarray}
in particular, 
 \begin{eqnarray}
 \label{p6}
 W_{1/2}(N,t) =\left(\frac{\sqrt{2\pi t}\,\, (et)^{-1/t}}{\Gamma\left(\frac{1}{t}\right)}
\right)^{N/2}\frac{{\prod_{l=1}^{N/2}\prod_{j=1}^{2l}(1-jt)}
}{ {\prod_{j=1}^{N/2}(1-2tj)}
}.
 \end{eqnarray}
 Therefore,
\begin{eqnarray}
\label{p7}
\frac{{\prod_{l=1}^{N/2}\prod_{j=1}^{2l}(1-jt)}
}{ {\prod_{j=1}^{N/2}(1-2tj)}
}&=&\prod_{j=0}^{N/2-1}\prod_{p=1}^{N-(2j+1)}(1-pt)\nonumber\\&=&\prod_{p=1}^{N/2}(1-(2p-1)t)^{N/2-p+1}(1-(2p)t)^{N/2-p},
 \end{eqnarray}
 where the last identity follows from $$ \prod_{p=1}^{N-(2j+1)}(1-pt)=\prod_{p=1}^{N/2-j}(1-(2p-1)t)\prod_{p=1}^{N/2-1-j}(1-2pt),$$ and $N$ is assumed to be even. As a result the partition  function $ W_{1/2}(N,t) $ may be written in terms of a single product as follows,
 \begin{eqnarray}
 \label{partition}
 W_{1/2}(N,t) =\left(\frac{\sqrt{2\pi t}\,\, (et)^{-1/t}}{\Gamma\left(\frac{1}{t}\right)}
\right)^{N/2}\prod_{p=1}^{N/2}(1-(2p-1)t)^{N/2-p+1}(1-(2p)t)^{N/2-p}.
 \end{eqnarray}
 Therefore, the free energy in this case reads
 \begin{eqnarray}
 \label{free energy}
2\log W_{1/2}(N,t) =\log\left(\left(\frac{\sqrt{2\pi t}\,\, (et)^{-1/t}}{\Gamma\left(\frac{1}{t}\right)}
\right)^{N}\prod_{p=1}^{N}(1-pt)^{N-p}\right)+\log\prod_{p=1}^{N/2}(1-(2p-1)t)
 \end{eqnarray}
 This is exactly the free energy obtained  for the  orthogonal Penner model in \cite{Chair1}, where the first term is the Penner free energy. This formula may be generalized as follows, first the above method, enables us to guess the following  general identity
 \begin{eqnarray}
\label{p8}
\frac{{\prod_{l=1}^K\prod_{j=1}^{ql}(1-jt)}
}{ {\prod_{j=1}^K(1-tqj)}}&=&\prod_{j=0}^{N/q-1}\prod_{p=1}^{N-(qj+1)}(1-pt)\nonumber\\&=&\prod_{p=1}^{N/q}(1-(qp-(q-1))t)^{N/q-p+1}(1-(qp-(q-2))t)^{N/q-p+1} \nonumber\\&&(1-(qp-(q-3))t)^{N/q-p+1}\cdots (1-(qp)t)^{N/q-p},
\end{eqnarray}
where the products on the the right hand side are  taken over non congruent and  congruent to $q$, of which   $q-1$ products are non congruent to $q$, and $N$ being a multiple of $q$. Therefore, the partition function connected with  the parametrized Euler characteristic $\xi^s_g(\gamma)$ is 
\begin{eqnarray}
\label{p9}
W_\frac{1}{q}(N,t)&&=\left(\frac{\sqrt{2\pi t}}{ \Gamma(\frac{1}{t})\,(et)^\frac{1}{t}}\right)^{N/q}\prod_{p=1}^{N/q}(1-(q(q-1))t)^{N/q-p+1}\nonumber\\&&(1-(qp-(q-2))t)^{N/q-p+1}(1-(qp-(q-3))t)^{N/q-p+1}\cdots\ (1-(qp)t)^{N/q-p}.\nonumber\\
\end{eqnarray}
Thus, the free energy $q\log W_\frac{1}{q}(N,t)$ that computes the parametrized Euler characteristic may be written as follows
\begin{eqnarray}
\label{p10}
q\log W_\frac{1}{q}(N,t) &=& N\log\left(\frac{\sqrt{2\pi t}}{ \Gamma(\frac{1}{t})\,(et)^\frac{1}{t}
}\right)\nonumber\\&+&\sum_{p=1}^{N/q}\left(N-(qp-(q-1)\right)\log\left(1-(qp-(q-1))t\right)\nonumber\\&+&\sum_{p=1}^{N/q}\left(N-(qp-(q-2)\right)\log\left(1-(qp-(q-2))t\right)\nonumber\\&+&\sum_{p=1}^{N/q}\left(1-(qp-(q-3)\right)\log\left(1-(qp-(q-3))t\right)\nonumber\\&+&\cdots+\sum_{p=1}^{N/q}\left(N-(qp-1)\right)\log\left(1-(qp-1))t\right)\nonumber\\&+&\sum_{p=1}^{N/q}\left(N-qp\right)\log\left(1-qpt\right)+\sum_{p=1}^{N/q}\log\left(1-(qp-(q-1))t\right)\nonumber\\&+&2\sum_{p=1}^{N/q}\log\left(1-(qp-(q-2))t\right)+\cdots+(q-1)\sum_{p=1}^{N/q}\log\left(1-(qp-1)t\right).\nonumber\\
\end{eqnarray}
Adding and subtracting the  terms $\sum_{p=1}^{N/q}\log\left(1-qpt\right) $ to the right hand of  Eq. (\ref{p10}), gives
\begin{eqnarray}
\label{p11}
q\log W_\frac{1}{q}(N,t)& =& \log\left(\frac{\sqrt{2\pi t}
}{ \Gamma(\frac{1}{t})\,(et)^\frac{1}{t}
}\right)^{N}+\sum_{p=1}^{N}\left(N-p\right)\log\left(1-pt\right)\nonumber\\&+&\sum_{p=1}^{N}\log\left(1-pt\right)-\sum_{p=1}^{N/q}\log\left(1-qpt\right)\nonumber\\&+&\sum_{p=1}^{N/q}\log\left(1-(qp-(q-2))t\right)\nonumber\\&+&2\sum_{p=1}^{N/q}\log\left(1-(qp-(q-3))t\right)+\cdots+(q-2)\sum_{p=1}^{N/q}\log\left(1-(qp-1)t\right),\nonumber\\.  
\end{eqnarray} 
In this formula, the first line is nothing but the the free energy of the Penner  model that we encountered previously for $q=2$ which computes the orbifold Euler characteristic of the moduli space of  Riemann surfaces  of genus $g$  with $s$ punctures $\chi(\cM_g^s)$ \cite{Harer2}, \cite{Penner}. Therefore, the parametrized Euler characteristic $\xi^s_g(\gamma)$ for any $q\geq2$, contains a contribution coming from the orbifold Euler characteristic of the moduli space of complex algebraic curves
given by 
\begin{equation}
\label{orb Euler}
\chi(\cM_g^s) = (-1)^s\frac{(g+s-2)!}{(g+1)(g-1)!}B_{g+1},
\end{equation}
 for   $g$ odd, here, $B_{g}$ is the $g$th Bernoulli number.  One should note that the third and fourth lines in Eq. (\ref{p11}) do contribute to  $\xi^s_g(\gamma)$ only for $q\geq3$, and the free energy given by Eq. (\ref{p11}) gives the well known results  for $q=1$, $q=2$. Next, we will derive a suitable  expression  for the free energy $ q\log W_\frac{1}{q}(N,t)$ that computes  $\xi^s_g(\gamma)$,  such that the first line is  omitted, since, this line  is the generating  function for  $\chi(\cM_g^s)$ . Thus, let $ q\log W_\frac{1}{q}^{1}(N,t)$  be this  contribution, and so, by expanding, the latter may be written as 
\begin{eqnarray}
\label{p12}
q\log W_\frac{1}{q}^{1}(N,t)&=&-\sum_{m\geq1}\frac{t^m}{m}\left(\sum_{p=1}^{N}p^{m}-\sum_{p=1}^{N/q}(qp)^{m}\right)-\sum_{m\geq1}\frac{t^m}{m} \sum_{j\geq0}^{m}(-1)^j\binom{m}{j}\sum_{p\geq1}^{N/q}q^{m}p^{m-j}\nonumber\\&\times&\left( (1-\frac{2}{q})^{j}+2(1-\frac{3}{q})^{j}+\cdots+(q-2)(\frac{1}{q})^{j}\right).
\end{eqnarray}
The power sum formula $$\sum_{j=1}^{n}j^k=\frac{1}{k+1}\sum_{r=1}^{k+1}\binom{k+1}{r}B_{k+1-r}
(-1)^{k+1-r}n^r, $$ may be be used to give
\begin{eqnarray}
\label{p13}
q\log W_\frac{1}{q}^{1}(N,t)&=&-\sum_{m\geq1}\frac{t^m}{m}\left(\frac{1}{m+1}\sum_{l=1}^{m+1}\binom{m+1}{l}(-1)^{m+1-l}B_{m+1-l}N^{l}\right)\nonumber\\&+&\sum_{m\geq1}\frac{t^m}{m}\left(\frac{q^m}{m+1}\sum_{l=1}^{m+1}\binom{m+1}{l}(-1)^{m+1-l}B_{m+1-l}q^{-l} N^{l}\right)\nonumber\\&-&\sum_{m\geq1}\frac{t^m}{m}\left(\sum_{j=0}^{m}(-1)^{j}\binom{m}{j}\frac{q^m}{m-j+1}\right)\nonumber\\&\times&\left(\sum_{l=1}^{m-j+1}\binom{m-j+1}{l}(-1)^{m-j+1-l}B_{m-j+1-l}q^{-l} N^{l}\right)\nonumber\\&\times&\left( (1-\frac{2}{q})^{j}+2(1-\frac{3}{q})^{j}+\cdots+(q-2)(\frac{1}{q})^{j}\right).
\end{eqnarray}
Let us now, extract  the coefficient of $s!(-1)^{s}N^{s}t^{g+s-1} $ in the expansion of $ q\log W_\frac{1}{q}^{1}(N,t)$, to that end,  we set $m=g+s-1$, $l=s$, then, we get
\begin{eqnarray}
\label{p14}
s!(-1)^{s}[N^{s}t^{g+s-1}]q\log W_\frac{1}{q}^{1}(N,t)&=&(-1)^{s+1}\frac{(g+s-2)!}{g!}(-1)^{g}\left(1-q^{g-1}\right)B_{g}\nonumber\\&+&(-1)^{s+1}\frac{(g+s-2)!}{g!}(-1)^{g}q^{g-1}\sum_{j=0}^{g}\binom{g}{j}B_{g-j}\nonumber\\&\times&\left( (1-\frac{2}{q})^{j}+2(1-\frac{3}{q})^{j}+\cdots+(q-2)(\frac{1}{q})^{j}\right).
\end{eqnarray}
For $g$  even, the first line given in Eq. (\ref{p11}) do not contribute to the parametrized Euler characteristic $\xi^s_g(\gamma)$, and hence the expression for $\xi^s_g(\gamma)$ reads
\begin{eqnarray}
\label{p15}
 \xi^s_g(\gamma)&=&(-1)^{s+1}\frac{(g+s-2)!}{g!}\left(1-q^{g-1}\right)B_{g}\nonumber\\&+&(-1)^{s+1}\frac{(g+s-2)!}{g!}q^{g-1}\left( B_{g}(\frac{2}{q})+2B_{g}(\frac{3}{q})+\cdots+(q-2)B_{g}(\frac{1}{q})\right), 
\end{eqnarray}
where we have used the formula for the Bernoulli polynomial in terms of the Bernoulli numbers $B_{n}(x)=\sum_{k=0}^{n}\binom{n}{k}B_{k}x^{n-k} $, and the property   $B_{n}(1-x)=(-1)^nB_{n}(x)$. By using the symmetry  $B_{g}(1-x)=B_{g}(x)$,  we may write $$ B_{g}(\frac{2}{q})+2B_{g}(\frac{3}{q})+\cdots+(q-2)B_{g}(\frac{1}{q})=\frac{q-2}{2}\left(2B_{g}(\frac{1}{q})+2B_{g}(\frac{2}{q})+\cdots+B_{g}(\frac{q/2}{q})\right), $$
now, the sum  on the right hand side  may be be written in a closed form by evaluating  the multiplication formula for the Bernoulli polynomials $$ B_{q}(kx)=k^{q-1}\sum_{k=0}^{q-1}\binom{n}{k}B_{q}(x+j/k), $$  at $x=0$, then, a simple computation shows$$2B_{g}(\frac{1}{q})+2B_{g}(\frac{2}{q})+\cdots+B_{g}(\frac{q/2}{q})=\left(\frac{1}{q^{g-1}}-1\right)B_{g}. $$ 
Therefore, if $g$ is even, the  parametrized Euler characteristic $\xi^s_g(\gamma)$ becomes
\begin{eqnarray}
\label{p16}
 \xi^s_g(\gamma)&=&(-1)^{s+1}\frac{(g+s-2)!}{g!}\left(1-q^{g-1}\right)B_{g}\nonumber\\&+&(-1)^{s+1}\frac{(g+s-2)!}{g!}q^{g-1}\left( \frac{q-2}{2} \right)\left(\frac{1}{q^{g-1}}-1\right)B_{g}\nonumber\\&=&(-1)^{s}\frac{(g+s-2)!}{g!2}\left(q^g-q\right)B_{g}.
\end{eqnarray}
This result is in complete agreement with the expression given by Goulden, Harer, and Jackson \cite{Harer1}. Let us now obtain the expression for  $\xi^s_g(\gamma)$ in the case that  $g$ is odd. This time, the contribution from the first line given in Eq. (\ref{p11}), corresponds to the orbifold Euler characteristic of the moduli space of complex algebraic curves  of genus $g$  with $s$ punctures $\chi(\cM_g^s)$. The other contributions for  $\xi^s_g(\gamma)$, comes from the last term of Eq. (\ref{p15}), since the first term does not contribute  for  $g$ odd. Then, if $g$ odd, 
\begin{eqnarray}
\label{p17}
 \xi^s_g(\gamma)&=&(-1)^s\frac{(g+s-2)!}{(g+1)(g-1)!}B_{g+1}\nonumber\\&+&(-1)^{s}\frac{(g+s-2)!}{g!}q^{g-1}\left( B_{g}(1-\frac{2}{q})+2B_{g}(1-\frac{3}{q})+\cdots+(q-2)B_{g}(\frac{1}{q})\right),
 \end{eqnarray}
 by using the symmetry $ B_{n}(1-x)=-B_{n}(x)$ for $n$ odd, then, the sum on the right hand  may be written as
 \begin{eqnarray}
\label{p18}
&&B_{g}(1-\frac{2}{q})+2B_{g}(1-\frac{3}{q})+\cdots+(q-2)B_{g}(\frac{1}{q})\nonumber\\&&=(q-2)B_{g}(\frac{1}{q})+(q-4)B_{g}(\frac{2}{q})+\cdots+B_{g}(\frac{q-1}{2q}), 
\end{eqnarray}
if $q$ is odd, and if $q$ is even the same formula is reached except that the last term is replaced by $ 2B_{g}(\frac{q-2}{2q})$. As a consequence, if $g$ is odd, the expression for the parametrized Euler characteristic becomes
\begin{eqnarray}
\label{p19}
 \xi^s_g(\gamma)&=&(-1)^s\frac{(g+s-2)!}{(g+1)(g-1)!}B_{g+1}\nonumber\\&+&(-1)^{s}\frac{(g+s-2)!}{g!}q^{g-1}\left(\sum_{i=1}^{(q-1)/2}(q-2i)B_{g}(\frac{i}{q})\right),\nonumber\\
 \end{eqnarray}
 for $q$ odd, while for $q$ even the maximum value of $i$ in the sum is $ (q-2)/2$. Therefore, the  parametrized Euler characteristic $ \xi^s_g(\gamma)$ of some   moduli spaces, as yet unidentified does contain the orbifold Euler characteristic of the moduli space of complex algebraic curves $\chi(\cM_g^s)$,  and other contributions that are linear combinations of the Bernoulli polynomials at rational arguments. Note that for $q=2$, that is, the real algebraic curves case, the parametrized Euler  characteristic is equal to  $\chi(\cM_g^s)$ for $g$ odd. As a result, for $g$ odd,  we expect the following  equality, 
\begin{eqnarray}
\label{p20}
 \xi^s_g(\gamma)&=&(-1)^s\frac{(g+s-2)!}{(g+1)(g-1)!}B_{g+1}\nonumber\\&+&(-1)^{s}\frac{(g+s-2)!}{g!}q^{g-1}\left(\sum_{i=1}^{(q-1)/2}(q-2i)B_{g}(\frac{i}{q})\right),\nonumber\\&=&\frac{{(g+s-2)!(-1)^{s+1}} }{ {(g+1)!}}
\left\{ (g+1)B_g q^g+
\sum_{r=0}^{g+1}\binom{ g+1}{r}B_{g+1-r}{B_{r}
}{q^{r}}\right\},
 \end{eqnarray}
 where the last expression in the above equation is  the parametrized Euler characteristic derived in \cite{Harer1}, for $g$ is odd.
 If $g=1$, then, the following formula is deduced
 \begin{eqnarray}
\label{p21}
\sum_{i=1}^{(q-1)/2}(q-2i)B_{1}(\frac{i}{q})=-\left(\frac{1}{12}q^{2}-\frac{1}{4}q+\frac{1}{6}\right),
\end{eqnarray}
while for odd $g>1$, one has
 \begin{eqnarray}
\label{p22}
\sum_{i=1}^{(q-1)/2}(q-2i)B_{g}(\frac{i}{q})=-q^{1-g}\left(B_{g+1}+\frac{1}{g+1}\sum_{r=1}^{g+1}\binom{ g+1}{r}B_{g+1-r}{B_{r}
}{q^{r}}\right),
\end{eqnarray}
from which the following interesting identity is obtained,
\begin{eqnarray}
\label{identity1}
\sum_{r=1}^{2g}\binom{ 2g}{r}B_{2g-r}B_{r}q^{r}&=&\sum_{r=1}^{g}\binom{ 2g}{2r}B_{2g-2r}B_{2r}q^{2r}\nonumber\\&=&(1-2g)B_{2g}-(2g)q^{2g-2}\sum_{i=1}^{(q-1)/2}(q-2i)B_{2g-1}(\frac{i}{q}).
\end{eqnarray}
Now, we know that the second sum given in Eq. (\ref{p19}), has contributions only for $q\geq 3$, therefore, we should have
\begin{eqnarray}
\label{identity2}
\sum_{r=1}^{2g}\binom{ 2g}{2r}B_{2g-2r}B_{2r}&=&\sum_{r=1}^{g}\binom{ 2g}{2r}B_{2g-2r}B_{2r}2^{2r}\nonumber\\&=&(1-2g)B_{2g}.
\end{eqnarray}
These are well known  formulae for Bernoulli numbers.
The consistency of the formulas given by Eq. (\ref{p21}), and Eq. (\ref{p22})  may be checked 
through the following simple  examples, the first formula for $q=3$, $q=4$ gives $ B_{1}(1/3)=-1/6$, $B_{1}(1/4)=-1/4$, respectively, 
and by 
setting $g=3$, $q=3$, $q=4$, then,  $B_{3}(1/3)= 1/27$, $B_{3}(1/4)= 3/64$, respectively, this is  in agreement with the direct evaluation of the Bernoulli polynomials at these rational values. It is interesting to note that if one uses the Almkvist-Meurman theorem \cite{Almkvist} which states that the poduct  $ q^gB_{g}(i/q)$ is an integer for $g$ odd ( $g>1$), and $ 0\leq i\leq q$, then,
\begin{equation} 
\label{integer}2 q^{g-1}\sum_{i=1}^{(q-1)/2}iB_{g}(\frac{i}{q})-\left(B_{g+1}+\frac{1}{g+1}\sum_{r=1}^{g+1}\binom{ g+1}{r}B_{g+1-r}{B_{r}
 }{q^{r}}\right),
 \end{equation} 
 must be an integer. We hope that the  method presented  here,  will be useful in giving a geometrical meaning to  these  moduli spaces, as yet unidentified and  whose  Euler characteristics are given by the  parametrized Euler characteristic  $ \xi^s_g(\gamma)$. Although, at present there is no interpretations of these moduli spaces, however, the free energy that computes the  parametrized Euler characteristic  $ \xi^s_g(\gamma)$  in the double-scaling limit reproduces the parametrized  Euler characteristic without punctures  and  share the same critical points as in the Penner model \cite{Chair2}, \cite{Vafa1}. To that end , let us write  the  free energy in the Harer-Goulden-Jackson model as
\begin{eqnarray}
\label{cont1}
F_{q}(N,t)&=\frac{1}{q}\sum_{g,s}\frac{(-1)^{s}}{s!}\xi^{s}_g(\gamma) N^{1-g}t^{g+s-1},
\end{eqnarray}
where we have used the natural scaling $ t\rightarrow t/N $. If $g$ is even, then, the  free energy reads
\begin{eqnarray}
\label{cont2}
F_{q}(N,t)&=&\sum_{g,s}\frac{(2g+s-2)!}{s!2}\left(q^{2g-1}-1\right)\frac{B_{2g}}{(2g)!}N^{1-2g}t^{2g-1+s}\nonumber\\&=&\frac{N}{2}\sum_{s=2}\frac{t^{s-1}}{s(s-1)}\left(\frac{1}{q}-1\right)\nonumber\\&+&\frac{1}{2}\sum_{g\geq 1}\sum_{s\geq 0}\frac{(2g+s-2)!}{s!}\left(q^{2g-1}-1\right)\frac{B_{2g}}{(2g)!}N^{1-2g}t^{2g-1+s}.
\end{eqnarray}
 The sum over punctures may be carried out to give 
 \begin{eqnarray}
\label{cont3}
F_{q}(N,t)&=&\frac{N}{2}\Bigr{(}1+(\frac{1-t}{t})\log(1-t)\Bigl{)}\left(\frac{1}{q}-1\right)\nonumber\\&+&\frac{1}{2}\sum_{g\geq 1}\Bigl(\frac{N(1-t)}{t}\Bigr)^{1-2g}\left(q^{2g-1}-1\right)\frac{B_{2g}}{(2g)(2g-1)},
\end{eqnarray}
to obtain the continuum limit of the free energy $F_{q}(N,t)$, set $ \mu=N(1-t)$ and let $N\rightarrow \infty$, $t \rightarrow 1$, such that $ \mu$ is kept fixed ("double scaling limit") to get
\begin{eqnarray}
\label{cont4}
F_{q}(\mu)&=&\frac{\mu}{2}\log\mu\left(\frac{1}{q}-1\right)\nonumber\\&+&\frac{1}{2}\sum_{g\geq 1}{\mu}^{1-2g}\left(q^{2g-1}-1\right)\frac{B_{2g}}{(2g)(2g-1)}.
\end{eqnarray}
Therefore, this  is the generalization of  the orthogonal Penner free energy in the continuum limit, that is, $q=2$ \cite{Chair1}, \cite{Chekhov} and having the same critical points as the Penner model \cite{Chair2}, \cite{Vafa1}.
If $g$ is odd, then, the free energy may be written as
\begin{eqnarray}
\label{cont5}
F_{q}(N,t)&=&\frac{1}{q}
\sum_{g,s}\frac{(2g+s-3)!}{s!(2g)!}(2g-1)B_{2g}N^{2-2g}t^{2g-2+s}\nonumber\\&+&\sum_{g,s}\frac{(2g+s-3)!}{s!(2g-1)!}q^{2g-3}\left(\sum_{i=1}^{(q-1)/2}(q-2i)B_{2g-1}(\frac{i}{q})\right)N^{2-2g}t^{2g-2+s},
 \end{eqnarray}
 where the first term is the free energy of the  Penner model discussed in a great details in \cite{Chair2}, while the second term is the  free energy contribution  for $q\geq 3$. The second term when summed over punctures gives
\begin{eqnarray}
\label{cont6}
&&\sum_{g,s}\frac{(2g+s-3)!}{s!(2g-1)!}q^{2g-3}\left(\sum_{i=1}^{(q-1)/2}(q-2i)B_{2g-1}(\frac{i}{q})\right)N^{2-2g}t^{2g-2+s}\nonumber\\&&=\log(1-t)\left( \frac{1}{12}q-\frac{1}{4}+\frac{1}{6q}\right)+\nonumber\\&&\sum_{g\geq 2}\Bigl(\frac{N(1-t)}{t}\Bigr)^{2-2g}\frac{q^{2g-3}}{(2g-1)(2g-2)}\left(\sum_{i=1}^{(q-1)/2}(q-2i)B_{2g-1}(\frac{i}{q})\right).\nonumber\\
\end{eqnarray}
 Finally, the continuum limit  reads
\begin{eqnarray}
\label{cont7}
F_{q}(\mu)&=&\frac{1}{2q}\mu^2 \log\mu-\frac{1}{12q}\log\mu+\frac{1}{q}\sum_{g\geq2}\frac{1}{(2g-2)}\frac{B_{2g}}{2g}\mu^{2-2g}\nonumber\\&+&\frac{1}{q}\left( \frac{1}{12}q^2-\frac{q}{4}+\frac{1}{6}\right)\log\mu\nonumber\\&+&\frac{1}{q}\sum_{g\geq 2}(\mu/q)^{2-2g}\frac{1}{(2g-1)(2g-2)}\left(\sum_{i=1}^{(q-1)/2}(q-2i)B_{2g-1}(\frac{i}{q})\right).
\end{eqnarray}
 If we set $q=2$, we recover our previous results \cite{Chair1} on the orthogonal Penner model in which the orientable contribution part gives half the Penner free energy. Note that for $q\geq3$, the  structure of the free energy is $1/q$ the orientable contribution plus  other contributions written in terms of the Bernoulli polynomial at rational argument. Here, in this sector  where $g$  is assumed to be odd, it may be possible to interpret the free energy  $ F_{q}(\mu)$ for $q\geq 3 $,  as some sort of corrections received by  the free energy for $q=2$. All genera receive  corrections except  the sphere. We have shown recently, that the continuum limit of both of $SO$ Chern-Simons gauge theory \cite{Vafa2} and $SO$ Penner  model are equivalent  \cite {Chair3}. Therefore, we may ask, if there is  a Chern-simons gauge theory whose free energy is the sum of  free energies given in Eq. (\ref{cont4})  and Eq. (\ref{cont7}), this may corresponds to a topological string on the quotient of the resolved conifold by the discrete group $Z_{q}$ . To conclude, in this work we gave an alternative formula  for the free energy that computes the  Parametrized Euler Characteristic of the Goulden-Harer-Jackson model. This formula, contains both the Penner and the orthogonal Penner free energies as special cases. Furthermore, for $g$, odd, the formula shows clearly, that the orbifold Euler characteristic of Riemann surfaces of genus $g$, with $s$  punctures, is always present in the expression for the parametrized Euler Charactristic for all the parameters $q$. This is also the case in the continuum limit, where the penner free energy in the continuum limit turns up in the total free energy of  Goulden-Harer-Jackson model in the continuum limit for all the parameters $q$. Therefore, this model is a deformation of the  Penner model such that the  critical points in the Penner model are fixed during  this deformation.
  
 \vspace{7mm} {\bf Acknowledgment:} I would like to thank G. Bonelli and  K.S. Narain for discussions and reading the manuscript. Also,I would  like to thank the ICTP, Trieste for the supports they give me.


\newpage 
\begin{thebibliography}{999}
\bibitem{Harer1}I.P. Goulden, J.L. Harer, and D.M. Jackson,
Trans. Amer. Math. Soc. \textbf{11} 353, 4405 (2001)
\bibitem{Harer2} J. Harer, D. Zagier,  Invent. math. \textbf{85},  457 (1986)
\bibitem{Penner} R. C. Penner  J.Diff. Geometry \text\bf {27} (1988), 35
\bibitem{Chekhov} L. Chekhov, A. Zabrodin, Mod. Phys. Lett. \textbf{A6}, 3143 (1991).
\bibitem{Chair1} M. Dalabeeh, N. Chair J. Phys. A: Math. Theor. \textbf{353} 465204 (2010)
\bibitem{Bessis}  D.Bessis, C.Itzykson and J.B.Zuber
Adv.\ Applied Math.
\textbf{1}, 109 (1980).
\bibitem{Almkvist} G.Almkvist and A. Meurman, C. R. Math. Rep. Acad. Sci. Canada.  \textbf{13}, 104 (1991).
\bibitem{Chair2} N.Chair Rev. Math. Phys. \text{3} 285 (1991)
\bibitem{Vafa1}J. Distler  and C.Vafa  Mod. Phys. Lett.\text{ A 6} 259 (1991)
\bibitem{Vafa2} S.Sinha, C.Vafa, hep-th 0012136
\bibitem{Chair3} N. Chair  M. Dalabeeh Progress of Theoretical Physics, \text{127}, No. 2, 179 (2012) 

\end{thebibliography}
\end{document}